\documentclass[apjl]{emulateapj}
\usepackage{natbib}
\usepackage{apjfonts}
\usepackage{epsfig}

\newcommand{\MUV}{M_\mathrm{UV}}
\newcommand{\zLP}{z_{\mathrm{AB}}}

\newcommand{\nbar}{\bar{n}}
\newcommand{\Nbar}{\bar{N}}
\newcommand{\Nfields}{N_\mathrm{fields}}
\newcommand{\Nreals}{N_\mathrm{reals}}
\newcommand{\sinc}{\mathrm{sinc}}
\newcommand{\rhobar}{\bar{\rho}}
\newcommand{\sigmaDM}{\sigma_{\mathrm{DM}}}
\newcommand{\sigmag}{\sigma_{\mathrm{g}}}
\newcommand{\sigmab}{\sigma_{b}}
\newcommand{\sigmaN}{\sigma_{\mathrm{N}}}
\newcommand{\deltaDM}{\delta_{\mathrm{DM}}}
\newcommand{\deltag}{\delta_{\mathrm{g}}}
\newcommand{\Deltab}{\Delta_{b}}
\newcommand{\phistar}{\phi_{\star}}
\newcommand{\Mstar}{M_{\star}}
\newcommand{\pLN}{p_\mathrm{LN}}
\newcommand{\bx}{\mathbf{x}}
\newcommand{\bk}{\mathbf{k}}

\newcommand{\ave}[1]{\langle{#1}\rangle}

\newcommand{\FSIXLP}{\mathrm{F600LP}}
\newcommand{\FNM}{\mathrm{F098M}}
\newcommand{\WMAP}{{\it Wilkinson Microwave Anisotropy Probe}}

\shorttitle{Measuring Bias from Cosmic Variance}
\shortauthors{Robertson}

\begin{document}

\title{A Method for Measuring the Bias \\ of High-Redshift Galaxies from Cosmic Variance}

\author{Brant E. Robertson\altaffilmark{1}}

\altaffiltext{1}{Hubble Fellow, brant@astro.caltech.edu}
\affil{Astronomy Department, California Institute of Technology, 
MC 249-17, 1200 East California Boulevard, Pasadena, CA 91125, USA}

\begin{abstract}
As deeper observations discover increasingly distant galaxies,
characterizing the properties of high-redshift galaxy populations will become increasingly challenging and paramount.
We present a method for measuring the clustering bias of high-redshift galaxies from the 
field-to-field scatter in their number densities induced by cosmic variance.  Multiple widely-separated
fields are observed to provide a large number of statistically-independent samples of the high-redshift
galaxy population.  
The expected Poisson uncertainty is removed from the measured dispersion in the distribution of
galaxy number counts observed across these many fields, leaving, on average, only the 
contribution to 
the scatter expected from cosmic variance.  With knowledge of 
the $\Lambda$-Cold Dark Matter power spectrum, the galaxy bias is then calculated from the measured cosmic 
variance.  The results of cosmological N-body simulations can then be used to estimate the halo mass
associated with the measured bias.  
We use Monte Carlo simulations
to demonstrate that Hubble Space Telescope pure parallel programs
will be able to determine galaxy 
bias at $z\gtrsim6$ using this method, 
complementing future measurements from correlation functions.
\end{abstract}


\section{Introduction}

Galaxy bias, the spatial clustering
strength of galaxies relative to the matter
field, encodes important information about
the connection between the luminous baryons
of galaxies and the dark matter (DM) halos 
that host them.  The galaxy bias $b$, defined as
the ratio between the
galaxy and matter correlation functions (CFs)
$b^{2}\equiv \xi_{gg}/\xi_{mm}$, 
reflects both the clustering of galaxy-scale DM
halos
and how luminous galaxies
populate halos of a given mass.
A self-consistent connection between
the luminosity function (LF), angular
CF, and 
mass-to-light ratio of galaxies can
be developed 
\citep[e.g.,][]{van_den_bosch2003a,vale2004a,conroy2006a,lee2009a}.

CF estimates of bias,
which require large samples of galaxies
spanning a wide range of angular
separations, 
are remarkably successful at low-redshift
\citep[e.g.,][]{norberg2001a,zehavi2002a}.
Angular CF
measurements on smaller samples
have been performed for Lyman 
Break Galaxies (LBGs)
\citep[][]{giavalisco1998a,arnouts1999a,ouchi2004a,ouchi2005a,foucaud2003a,hamana2004a,adelberger2005a,allen2005a,kashikawa2006a,lee2006a,lee2009a}
and Lyman-$\alpha$ emitters
\citep{ouchi2003a,hamana2004a,kovac2007a,gawiser2007a}
over a range of redshifts.
These studies have extended 
high-significance $(>3\sigma)$ detections of 
bias out to $z\sim5$.

At the highest redshifts ($z\gtrsim5.5$), 
limited samples have made 
significant detections of galaxy bias 
difficult.  \citet[][]{mclure2009a} used a
$0.63$deg$^{2}$ area observed by the 
UKIRT Infrared Deep Sky Survey Ultra Deep
Survey and the Subaru XMM-Newton Survey
to measure the clustering of $5<z<6$ LBGs
with $z<26$AB ($b=5.2_{-0.8}^{+1.2}$), but
most of their systems lie at redshifts $z<5.5$.
\citet[][]{overzier2006a}
reported a measurement of bias
for $i$-dropout galaxies in Great Observatories 
Origins Deep Survey (GOODS)
($b=4.1_{-2.6}^{+1.5}$) but no
detection of 
clustering in the Ultra Deep Field.
A study of $z$-dropout
galaxies in the Subaru Deep Field
and GOODS-N by \citet{ouchi2009a}
indicated that $z\sim7$ galaxies may be strongly clustered, 
but measuring bias from
only 22 candidates 
is challenging.

Given the difficulty of measuring clustering
of high-redshift galaxies from the angular
CF, an exploration of 
complementary approaches is warranted. 
As a salient example,
\citet{adelberger1998a} used a
counts-in-cells analysis 
\citep[][]{peebles1980a} 
to measure the clustering of
redshift $z\sim3$ LBGs from 
statistical fluctuations
in their number densities.
Having obtained 
spectroscopic redshifts for $N\sim270$
LBGs in six fields, they 
split their samples into redshift slices
and calculated the variance in galaxy
counts in rectangular volumes.
The bias was then
measured from the excess dispersion in the
number counts beyond the Poisson
variance.

In this {\it Letter}, we present calculations
that demonstrate the bias of high-redshift
($z\gtrsim6$) galaxies can also be measured from
field-to-field dispersion in galaxy counts
from photometric dropout samples.  
Specifically, we use Monte Carlo simulations to demonstrate that
pure parallel programs with the Hubble Space
Telescope (HST) that 
obtain large numbers ($\gtrsim100$)
of statistically-independent samples of $z\approx6-7$
galaxies with the
Advanced Camera for Surveys (ACS) 
or Wide Field Camera 3 (WFC3) 
\citep[e.g.,][]{ratnatunga2002a,sparks2002a,trenti2008b,yan2008a} 
can provide a significant measurement of
high-redshift galaxy bias.
Throughout, we adopt the cosmological parameter values
determined by the $5$-year \WMAP~joint analysis \citep{komatsu2009a}.

\section{Measuring Halo Bias from Cosmic Variance}

Consider a number $\Nfields$ of 
telescope pointings on the sky.
At the location $\bx_{i}$ of the $i$-th field
the expected number density of galaxies
is
\begin{equation}
\label{eqn:number_counts}
n(\bx_{i}) = \nbar [ 1 + b\deltaDM(\bx_{i})],
\end{equation}
\noindent
where $\nbar$ is the mean average galaxy number density,
$b$ is the bias,
and $\deltaDM\equiv (\rho-\rhobar)/\rhobar)$ 
is the matter overdensity 
of the field.
The matter variance $\ave{\deltaDM^{2}}$, averaged
over many survey volumes $V$, is
\begin{equation}
\label{eqn:dm_variance}
\sigmaDM^{2}(V,z) \equiv \ave{\deltaDM^{2}} = D^{2}(z) \int \frac{d^{3}k}{(2\pi)^{3}} P(k) |\hat{W}(\bk,V)|^{2}
\end{equation}
\noindent
where $D(z)$ is the growth factor at
redshift $z$, $P(k)$ is the matter
power spectrum (we use the \citealt{eisenstein1998a} transfer function), and 
\begin{equation}
\hat{W}(\bk,V) = \sinc \left( \frac{k_{x}r\Theta_{x}}{2} \right) \sinc \left( \frac{k_{y}r\Theta_{y}}{2} \right) \sinc \left( \frac{k_{z}\delta r}{2} \right)
\end{equation}
\noindent
is the Fourier transform of the rectangular volume 
$V = r\Theta_{x} \times r\Theta_{y} \times \delta r$ with
angular area $\Theta_{x} \times \Theta_{y}$ and depth $\delta r$
at distance $r$.  The expected dispersion in
galaxy counts owing to cosmic variance is simply 
$\sigmag^{2}=b^{2}\sigmaDM^{2}$
\citep[e.g.,][]{stark2007a,trenti2008a,munoz2010a,robertson2010a}.

With a sample of $\Nfields$ fields, the
distribution of galaxy counts $N$,
with mean $\Nbar$ per field, will have a dispersion
$\sigmaN^{2}\equiv\ave{(N-\Nbar)^{2}}\approx\sigmag^{2}\Nbar^{2} + \Nbar$
that includes contributions from both cosmic
and Poisson variance.  The bias
can be estimated as
\begin{equation}
\label{eqn:bias_measure}
b^{2} \approx \frac{\sigmaN^{2}-\Nbar}{\Nbar^{2}\sigmaDM^{2}}.
\end{equation}
\noindent
Below, we demonstrate that this estimate can be used to measure
the bias of high-redshift galaxies 
from multiple widely-separated HST pointings.
The accuracy of this method will depend on
$\Nfields$ and the relative size of the
cosmic and Poisson variances, and we now explore these effects
through Monte Carlo simulations.

\section{Monte Carlo Simulations of Galaxy Bias Measurements from Cosmic Variance}
\label{section:bias_measurement}

To simulate high-redshift galaxy
number counts, we require a model for the distribution
of galaxy counts for each field
in our sample.  Below, we describe our model distribution
and our Monte
Carlo realizations of this distribution that act
as our model observational samples.

The initial cosmological linear overdensity field is a Gaussian
random field with dispersion $\sigmaDM^{2}$ given
by Equation \ref{eqn:dm_variance}.  As the density
field evolves owing to gravitation, 
the overdensity distribution changes to maintain
the positivity condition on the density 
($\deltaDM\ge-1$).  
The shape of the quasi-linear
overdensity distribution can be
approximated by a
lognormal \citep{coles1991a,kofman1994a}.
The corresponding
galaxy overdensity distribution will be
lognormal, written as
\begin{equation}
\label{eqn:galaxy_overdensity_pdf}
\pLN(\deltag|\sigmag^{2})  = \frac{1}{\sqrt{2\pi x^{2}}}\exp\left[-\frac{1}{2}\left(\frac{y}{x} + \frac{x}{2}\right)^{2}\right],
\end{equation}
\noindent
where $y=(1+\deltag)$ and $x=(1+\sigmag^{2})^{1/2}$.
The expected cosmic variance is $\sigmag<1$ (see below),
so the quasi-linear distribution is appropriate.

At each location $\bx_{i}$, the number of galaxies
observed will be a Poisson-sampled random
variate of 
Equation \ref{eqn:galaxy_overdensity_pdf}.  The
probability of observing $N$ galaxies 
in fields of volume $V$
with cosmic variance $\sigmag^{2}$,
given the expected number $\Nbar =\nbar V$,
is
\begin{equation}
\label{eqn:galaxy_count_pdf}
p(N|\Nbar,\sigmag^{2}) = \frac{1}{N!}\int_{-1}^{\infty}\mathrm{d}\deltag \pLN(\deltag|\sigmag^{2})[(1+\deltag)\Nbar]^{N}e^{-(1+\deltag)\Nbar}
\end{equation}
\noindent
\citep[][]{adelberger1998a}.  
Our simulations of the bias measurement
consist of drawing discrete random samples from this 
distribution, as described below.

\subsection{Observational Model}
\label{subsection:observational_model}

Measuring the bias from the dispersion in 
galaxy counts requires an observational
program suited to probing many
high-redshift galaxy samples.
Our presented method is motivated by 
HST programs 
(Programs 9488, PI Ratnatunga, 9575 and 9584, PI Sparks, 11700, PI Trenti, and 11702, PI Yan)
that use pure parallel
observations with ACS or WFC3
to probe $z\gtrsim6-7$ galaxy populations.  
While the total numbers of
long-duration and multiple-passband observations in
the HST archive are few ($\sim10s$),
we expect that similar pure parallel
observations will be obtained
by HST in ongoing and future cycles.
We will therefore model future HST pure parallel
programs that obtain ACS $i$- and $z$-band data, which would allow
for $i$-dropout selection at $z\sim6$ to $z\sim26.5$AB
\citep[single orbit exposure, extrapolated from][]{beckwith2006a}.  
Additionally, as an example comparison with WFC3 
observations, we will model the
Yan WFC3 program that can select $\FSIXLP-\FNM$
dropouts at $6.5\lesssim z\lesssim7.3$, and 
reach $\FNM\sim27$AB magnitude sensitivity (single orbit exposure),
but our method can be adapted for use with any WFC3 filter strategy.  
We adopt these
survey designs, summarized in Table \ref{table:surveys}, 
as our points of comparison for modeling 
observational measurements of high-redshift galaxy bias.

We wish to simulate the number of high-redshift galaxies
these pure parallel surveys will discover in each field.
The
mean 
abundance $\Nbar$ per field
is determined by
integrating a \citet{schechter1976a} LF
over the volume probed by each pointing.
For modeling the ACS observations of $i$-dropouts at $5.5\le z \le 6.5$, 
we adopt the measured $z\sim6$ LF parameters
$\phistar=0.0014~\mathrm{Mpc}^{-3} \mathrm{mag}^{-1}$,
$\Mstar=-20.24$AB, and $\alpha=-1.74$
\citep{bouwens2007a}.  With these values, we
expect
$\Nbar=4.3$ per $11.3$ arcmin$^{2}$ ACS field
in the magnitude range $\MUV\lesssim-20.5$.
To approximate the conversion between
$\zLP$ apparent and restframe $\MUV$
used by 
\citet{bouwens2006a}, we use 
$\MUV-\zLP \approx -47$.

For modeling the WFC3 observations 
the 
$z\sim7$ LF is uncertain at $6.5\le z \le 7.3$,  
but if we adopt the measured $z\sim7$ LF parameters
$\phistar=0.0014~\mathrm{Mpc}^{-3} \mathrm{mag}^{-1}$,
$\Mstar=-19.91$AB, and $\alpha=-1.77$
\citep{oesch2010a}, 
we expect $\Nbar=1.04$ per  $4.7$ arcmin$^{2}$ WFC3
field in the UV absolute magnitude range $\MUV\lesssim-20.2$.  
Our results will
be similar for other estimates of the $z\sim7$
LF \citep[e.g.,][]{mclure2010a,yan2009a}.

Galaxy counts for each field can be
generated by discretely sampling the 
Equation 
\ref{eqn:galaxy_count_pdf}, with
a variance $\sigmag$ and
bias $b$ determined
using the 
abundance-matching
methodology presented in 
\citet{robertson2010a}. This method
assumes the galaxies occupy 
DM halos with an abundance and bias
appropriate for their high redshifts 
\citep{tinker2008a,tinker2010a}.

\begin{deluxetable*}{lcccccccccc}
\tablecolumns{11}
\tablewidth{0pc}
\tablecaption{Model Parallel Surveys\label{table:surveys}}
\tablehead{ \colhead{Redshift} & \colhead{Detector} & \colhead{Orbits/Field\tablenotemark{a}} & \colhead{Depth} & \colhead{$\phistar$} & \colhead{$\Mstar$} & \colhead{$\alpha$} & \colhead{$\Nbar$\tablenotemark{b}} & \colhead{$\sigma_{\mathrm{g}}$} & \colhead{$1/\sqrt{\Nbar}$} & \colhead{Fields/$SNR\sim3$\tablenotemark{c}} \\
& & & \colhead{[AB mag]} & \colhead{[Mpc$^{-3}$ mag$^{-1}$]} & \colhead{[AB mag]} & & & &
}
\startdata
$5.5\leq z\leq 6.5$ & ACS  & 4 & $z\sim26.5$ & $1.4\times10^{-3}\,$\tablenotemark{d} & -20.24\tablenotemark{d} & -1.74\tablenotemark{d} & 4.29 & 0.37 & 0.48 & 40\\
$6.5\leq z\leq 7.3$ & WFC3 & 4 & $Y\sim27$ & $1.4\times10^{-3}\,$\tablenotemark{e} & -19.91\tablenotemark{e} & -1.77\tablenotemark{e} & 1.04 & 0.50 & 0.98 & 130
\enddata
\tablenotetext{a}{Includes single orbits for the detection, dropout, and bluer veto bands.}
\tablenotetext{b}{Average number of galaxies per field.}
\tablenotetext{c}{Number of fields required to reach $SNR\sim3$.}
\tablenotetext{d}{Adopted from \citet{bouwens2007a}.}
\tablenotetext{e}{Adopted from \citet{oesch2010a}.} 
\end{deluxetable*}

\begin{figure*}
\figurenum{1}
\epsscale{1}
\plottwo{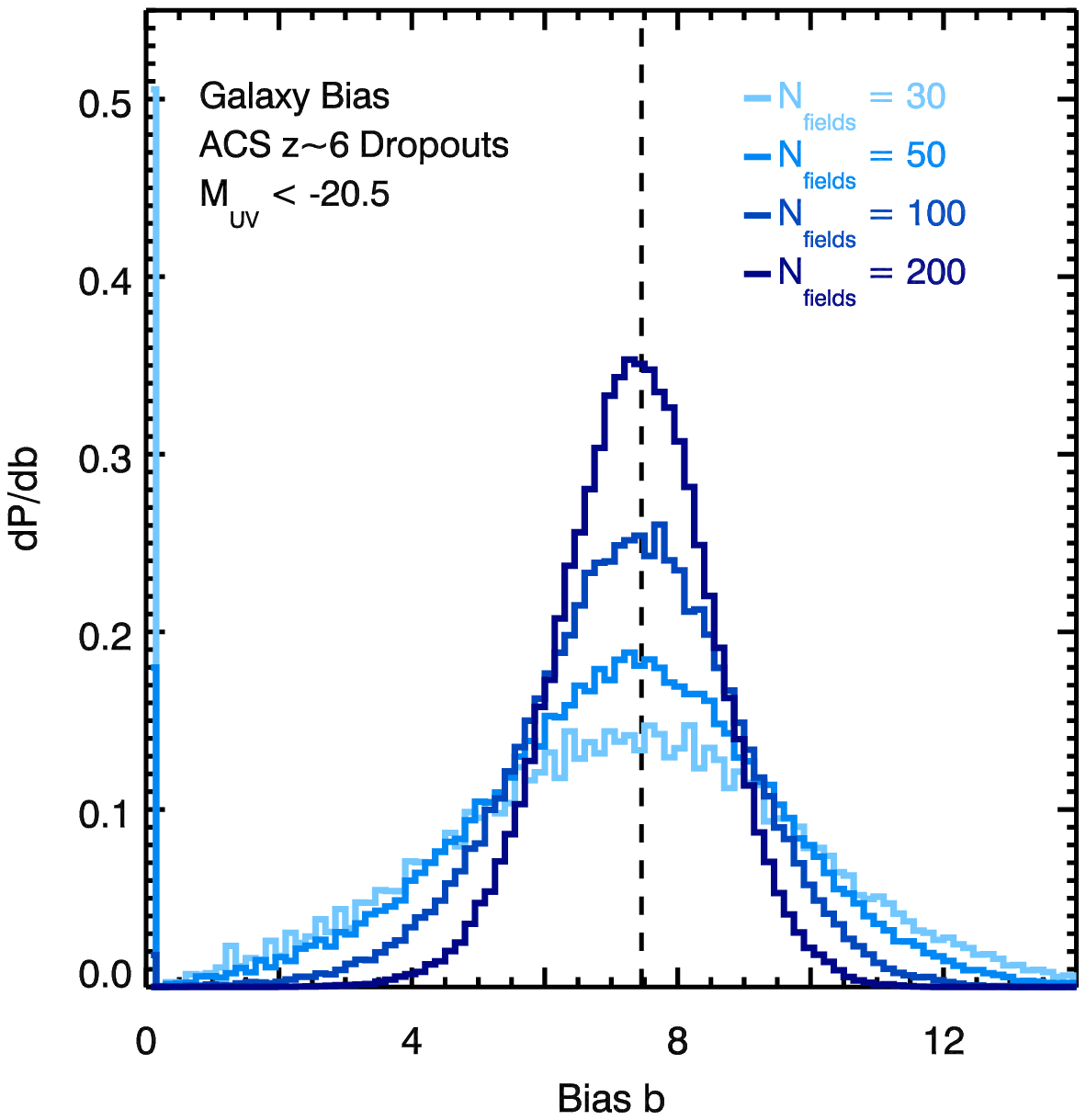}{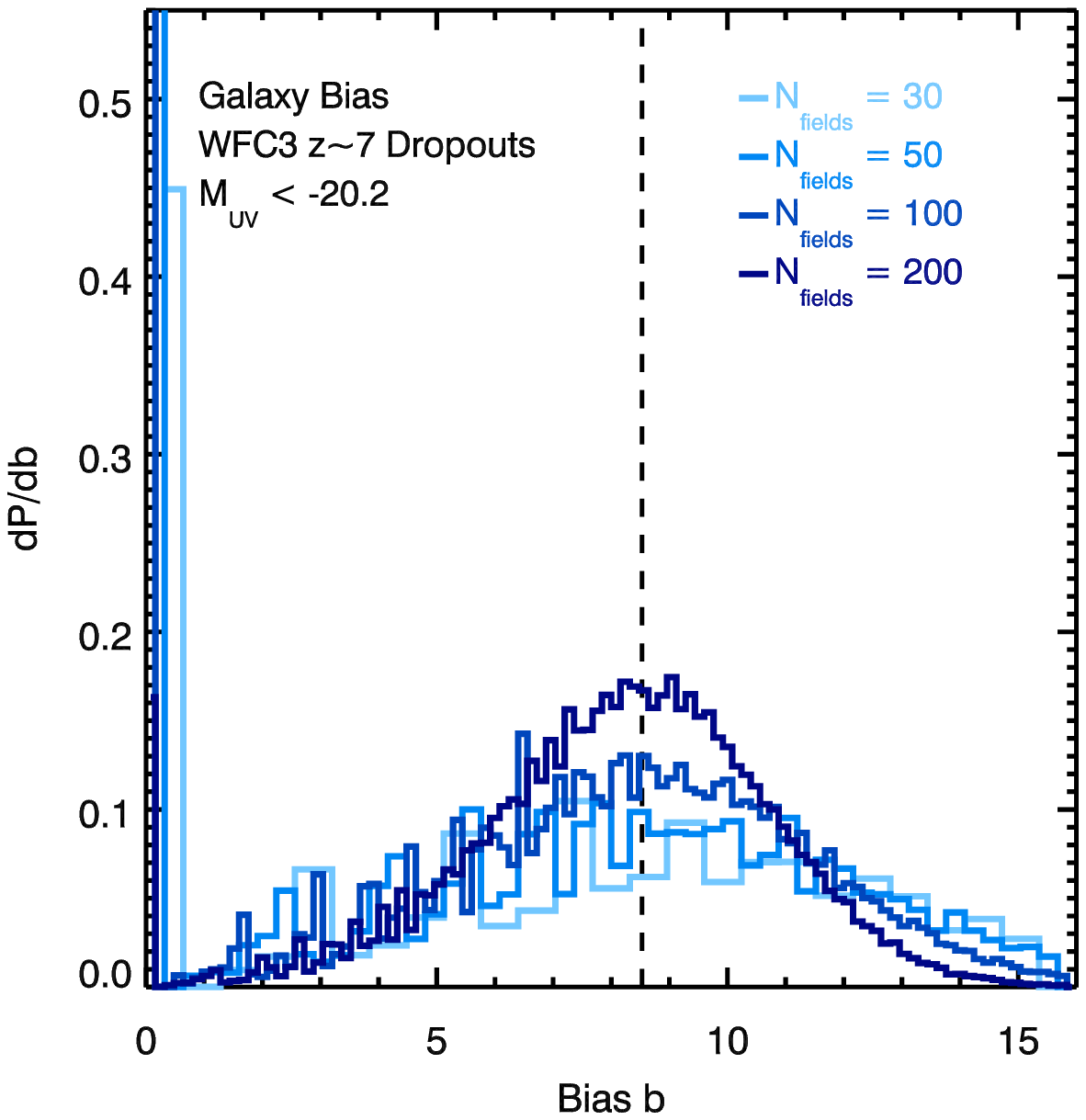}
\caption{\label{fig:bias_constraint}
Monte Carlo simulations of galaxy bias measurements from the field-to-field
variation in number counts across
$N=30-200$ (light to dark blue) statistically-independent, 
widely-separated fields.
Shown are 100,000 realizations of the measurement for 
$z\sim6$ ACS $i$-band (left panel) and
$z\sim7$ WFC3 $\FSIXLP$-band (right panel)
dropouts with UV absolute magnitudes $\MUV\lesssim-20.5$AB and $\MUV\lesssim-20.2$AB, respectively.
As the number of independent fields increases, the 
$SNR$ of the cosmic variance measurement increases rapidly.  
For some realizations the 
scatter in the distribution of number counts is smaller than Poisson and 
cosmic variance cannot be measured ($b=0$ bins, for a discussion see \S \ref{subsection:uncorrelated_acs_fields}).
}
\end{figure*} 

\subsection{Uncorrelated ACS Fields}
\label{subsection:uncorrelated_acs_fields}

The left panel of Figure \ref{fig:bias_constraint}
shows $\Nreals=100,000$ realizations of the galaxy
bias 
measurement for $z\sim6$ galaxies with 
$\MUV\lesssim-20.5$AB
across $\Nfields=30-200$ (light to dark blue)
widely-separated
ACS fields.  For this example
magnitude range at redshifts $5.5\leq z\leq6.5$, the mean 
number of galaxies per 11.3 arcmin$^{2}$ is $\Nbar=4.3$ and $\sigmaDM\approx0.05$.
The model bias calculated
for this galaxy abundance is 
$b\approx7.5$
(vertical dashed line).
The mean galaxy count per ACS field $\Nbar$ and 
the sample variance $\sigmaN^{2}$ are
calculated from each realization.
The ``observed'' bias is then estimated
using Equation \ref{eqn:bias_measure},
histogrammed, and
normalized such that the area under each curve is unity.

For each $\Nfields$-sized sample the mean expected 
measured bias is accurate,
with $\Deltab\equiv(\ave{b}-b)/b \approx -0.025$ for
$\Nfields=30$ decreasing to $\Deltab \approx -0.014$ for
$\Nfields=200$.
The expected uncertainty $\sigmab$ in the
measured bias decreases
from $\sigmab/b\approx0.36$ for $\Nfields=30$ to $\sigmab/b\approx0.16$
for $\Nfields=200$, improving as $\propto1/\sqrt{\Nfields}$.  
When $\Nfields$ is small, the measured dispersion
$\sigmaN^{2}$ can be less than $\Nbar$,  providing $b^{2}<0$
in Equation \ref{eqn:bias_measure}.  Such catastrophic failures happen infrequently
for ACS $i$-dropout samples
($<7.5\%$ for $\Nfields=30$ and $<0.3\%$ for $\Nfields=100$,
shown as the $b=0$ bin).

Given these results, we suggest that using the dispersion in 
galaxy counts determined from ACS photometric 
samples in uncorrelated fields to measure bias will be a
novel method for learning about the spatial clustering
of $z\sim6$ galaxies.  The bias measurement significance
increases from $SNR\approx3$ with $\Nfields=40$ to $SNR\approx6.4$ with 
$\Nfields=200$ fields for $i$-dropouts with magnitudes near $\Mstar$, 
providing a method for measuring
bias that increases in sensitivity with an 
increasing number of independent samples.

\subsection{Uncorrelated WFC3 Fields}
\label{subsection:uncorrelated_wfc3_fields}

The right panel of Figure \ref{fig:bias_constraint}
shows $\Nreals=100,000$ realizations of the 
bias 
measurement for $z\sim7$ galaxies with absolute
magnitudes
$\MUV\lesssim-20.2$AB
across $\Nfields=30-200$ (light to dark blue)
widely-separated
WFC3 fields.  For this 
magnitude range at redshifts $6.5\leq z\leq7.3$, the expected mean 
number of galaxies per 4.7 arcmin$^{2}$ is $\Nbar=1.04$ and $\sigmaDM\approx0.06$.
The model bias calculated
for this galaxy abundance is 
$b\approx8.5$ (vertical dashed line).
The observed mean galaxy count per WFC3 field $\Nbar$ and 
the sample variance $\sigmaN^{2}$ are
calculated from each of the $\Nreals$ realizations.
The ``observed'' bias is then estimated
using Equation \ref{eqn:bias_measure},
histogrammed, and
normalized such that the area under each curve is unity.

These histograms demonstrate that characterizing
the bias of rare
$z\sim7$ galaxies will be challenging.
Approximately $\Nfields=130$ pointings
are needed to measure the bias of these galaxies with an appreciable signal-to-noise 
($SNR\sim3$) in an accurate way ($\Deltab\lesssim -0.023$).
With fewer fields ($\Nfields\le100$), the expected
uncertainty in the measured bias is large 
($\sigmab/b\ge0.36$) and overestimated ($\Deltab\gtrsim 0.05$).
When $\Nfields$ is small, the measured dispersion
$\sigmaN^{2}$ can be less than $\Nbar$,  providing $b^{2}<0$
in Equation \ref{eqn:bias_measure}.  In these catastrophic failures
($26\%$ for $\Nfields=30$ and $9\%$ for $\Nfields=100$,
shown as the $b=0$ bin),
an unusually small Poisson scatter limits sensitivity to the
cosmic variance in number counts.  These failures decrease
as the number of pointings increases (to $3\%$ for $\Nfields=200$).

The larger numbers
of $z\sim6$ galaxies per ACS field compared with the number of
$z\sim7$ galaxies per WFC3 field (see 
\S \ref{subsection:uncorrelated_acs_fields}) makes the
$z\sim6$ ACS experiment much easier.
While $\Nfields\sim130$ pointings are required to 
significantly ($SNR\sim3$) measure the bias, such
a determination would provide a crucial constraint on
the nature of $z\sim7$ galaxies.  Current
HST programs may have too few fields to perform
this experiment, but extensions or new pure parallel 
programs will likely provide enough pointings in the
next few HST cycles to successfully measure the bias of
$z\sim7$ from cosmic variance.

\subsection{Correlated Fields}
\label{subsection:correlated_fields}

Pure parallel observations with ACS and WFC3
provide a novel method to obtain
statistically-independent samples of the
high-redshift galaxy population. However,
the forthcoming HST Multi-Cycle Treasury (MCT)
programs will probe greater numbers of contiguous pointings.
While the MCT areas
may be used to measure the angular correlation
function of high-redshift galaxies, one may wonder
whether contiguous areas may be used to measure
bias from cosmic variance.  Here, we
estimate the effects of
correlations in the galaxy counts induced by
large-scale density modes shared between nearby
pointings.  We consider $z\sim6$ dropouts in ACS 
pointings, as the improved statistics relative to
$z\sim7$ dropouts more clearly
illustrate the effects of field-to-field correlations.

The detailed correlation between nearby fields will 
depend on the geometry of the tiling and
the separation between fields, and
are difficult to calculate for general
survey designs.  However, we estimate 
the field-to-field correlations in
a contiguous tiling
as follows 
\citep[for another estimate, see][]{hu2006a}.  

Consider as before $\Nfields$ pointings,
but arranged in a contiguous tiling.  The cosmic
variance of an independent field with volume $V$ 
will depend on $\sigmag^{2}(V,z)$.  Correspondingly,
the galaxy counts in the larger volume 
$\Nfields V$ will have a sample variance 
$\sigmag^{2}(\Nfields V,z)$.  The number counts within
the $i$-th field of the $\Nfields$ pointings within the contiguous 
survey will then be correlated with the other $j$ fields 
approximately at the level of
\begin{equation}
\label{eqn:correlation_estimate}
\rho \equiv \frac{\sigma_{ij}^{2}}{\sigma_{ii}\sigma_{jj}}\approx \frac{\sigmag^{2}(\Nfields V,z)}{\sigmag^{2}(V,z)}.
\end{equation}
\noindent
This typically underestimates the
correlation between very nearby or adjacent fields,
but should approximate the average correlation of the
most widely-spaced pointings within the contiguous
area.  
The matrix describing the correlation between fields
$i$ and $j$ then has a simple form, 
with the diagonal elements $\Sigma_{ii}=1$
and off-diagonal elements $\Sigma_{i,j\ne i}=\rho$.

\begin{figure}
\figurenum{2}
\epsscale{1}
\plotone{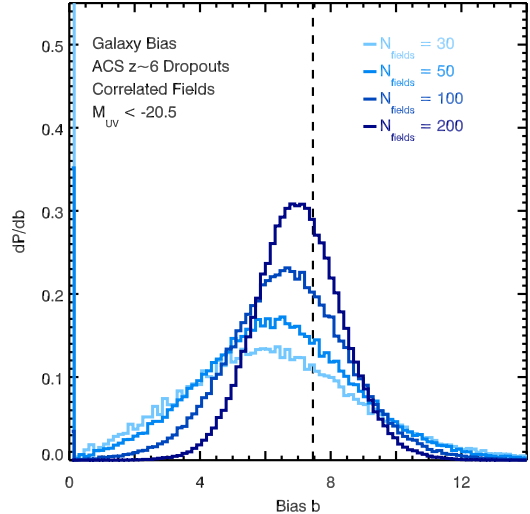}
\caption{\label{fig:bias_constraint_correlated}
Same as left panel of Figure 1, but for a contiguous survey where
large scale structure correlates 
the number counts in each field.
The field-to-field correlations
decrease the dispersion in galaxy counts,
and leads to a 10-20\% underestimate of the galaxy bias.
}
\end{figure}

To model the counts in correlated fields,
we use a standard Cholesky decomposition technique
to instill the
distribution of galaxy overdensities
with the correlation matrix $\Sigma$,
and then
Poisson-sample the correlated distribution to
generate the counts in each field.
Figure \ref{fig:bias_constraint_correlated} shows the
results of this procedure.
Since the number counts in
fields are correlated ($\rho\approx0.3$ for $\Nfields=30$ and $\rho\approx0.09$ 
for $\Nfields=200$), the dispersion $\sigmaN^{2}$ in the
ACS galaxy counts becomes smaller than in the uncorrelated case.
The resulting estimates for the cosmic variance are
correspondingly smaller, leading to a bias measurement that
is underestimated ($\Deltab =-0.15$ for $\Nfields=30$ 
and $\Deltab=-0.06$ for
$\Nfields=200$) and is more susceptible to catastrophic (measured $b\le0$) failures 
($13.6\%$ for $\Nfields=30$).

\subsection{Estimating Halo Masses from Galaxy Bias}
\label{subsection:halo_mass}

The simulations presented in \S
\ref{subsection:uncorrelated_acs_fields} and
\ref{subsection:uncorrelated_wfc3_fields}
demonstrate that the bias of high-redshift galaxies can 
be measured directly from the variance in number counts
between uncorrelated fields.  Using the connection between
bias and DM halo mass calculated from cosmological
simulations \citep{tinker2008a,tinker2010a}, the galaxy bias
can be used to estimate the characteristic mass of
DM halos with similar spatial clustering strength.
For detailed discussions of the connection between
halo mass and bias, we refer the reader to 
\cite{robertson2010a} and \cite{tinker2010a}.

\begin{figure}
\figurenum{3}
\epsscale{1}
\plotone{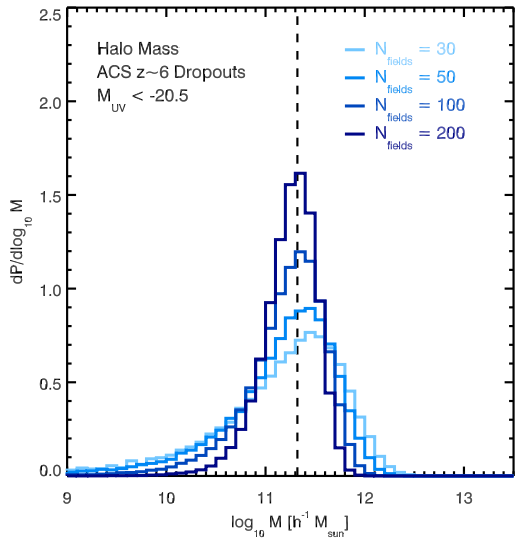}
\caption{\label{fig:mass_constraint}
Characteristic halo mass inferred from the
bias of $z\sim6$ ACS $i$-band 
dropouts with UV absolute magnitudes $\MUV\lesssim-20.5$AB. 
The bias is measured from number count
variations across $N=30-200$ (light to dark blue)
fields (see left panel of Fig \ref{fig:bias_constraint}),
and then converted into a mass
using the \citet[][]{tinker2010a} bias function for DM
halos.  
}
\end{figure}

Figure \ref{fig:mass_constraint}
shows the characteristic halo mass estimated by converting the bias 
(Figure \ref{fig:bias_constraint})
using the \cite{tinker2010a} bias-mass relation determined
from cosmological simulations.
For the case of $\MUV\lesssim-20.5$AB magnitude
$i$-dropout galaxies at redshift $z\sim6$ with bias $b\approx7.5$, the
mass of halos with the same bias at $z\sim6$ is
$M_{\mathrm{halo}}\approx2.1\times10^{11} h^{-1}M_{\sun}$ (dashed line).  
The bias can provide a 
precise estimate of a characteristic DM halo mass ($8\%$ in $\log M$ for $\Nfields=30$ and
$3\%$ in $\log M$ for $\Nfields=200$) that, while skewed, is accurate
($\Delta_{M}\equiv[\ave{\log M}-\log M]/\log M = -0.03$ for $\Nfields=30$ and 
$\Delta_{M}= -0.007$ for $\Nfields=200$).
This measure of a characteristic halo mass
only corresponds to the halo mass of
$i$-dropout galaxies if each halo hosts
one galaxy and mass strongly correlates 
with luminosity 
\citep[see][]{robertson2010a}, but
provides a convenient 
conceptual tool for discussing the approximate mass scale of
high-redshift galaxies.

\section{Summary and Discussion}

We have presented a simple counts-in-cells
method \citep[e.g.,][]{peebles1980a,adelberger1998a} for measuring the bias of
high-redshift galaxies from the field-to-field
variation in number counts induced
by cosmic variance.  The number of
high-redshift galaxies of a given luminosity are measured in
a large number independent, widely-separated
fields.  The Poisson contribution to the variance in 
the number counts across these fields is removed,
leaving a cosmic variance contribution
that depends on the bias and the mean
matter overdensity fluctuations on the scale of
the survey field volume.
We use Monte Carlo simulations of the
bias measurement to estimate the effectiveness of the
method for example galaxy populations, 
$i$-dropouts
at $z\sim6$ in ACS-sized fields 
and
$\FSIXLP$-dropouts 
at $z\sim7$ in
WFC3-sized fields.
Our method can be adapted for other dropout filter 
strategies.
At $z\sim6$ ($z\sim7$)
the method should provide a measurement
of galaxy bias with $SNR\gtrsim3$
for luminosities near $\MUV\approx\Mstar$ 
when 
$\Nfields\ge40$ ($\Nfields\gtrsim130$) fields
are used.
The uncertainty on the galaxy bias 
improves with an increasing number of independent
fields as $\approx1/\sqrt{\Nfields}$.
These requirements on the number of fields are 
concomitant with
the expected number of pointings available from
extensions to 
ongoing pure parallel HST programs 
\citep[][]{trenti2008b,yan2008a}, and the potential
for using this method to measure high-redshift galaxy
bias is therefore promising.
We also 
show that using correlated, nearby fields
to perform the presented measurement 
typically leads to an underestimate of the bias
owing to a correlation-induced decrease in 
the dispersion of galaxy
number counts.

The measured bias can be associated with 
a characteristic mass of DM halos with
similar clustering.  We show that the typical
uncertainty in the bias measured using the
presented method corresponds to an uncertainty
in the inferred DM halo mass of $\sim$
a few percent in $\log M$ for $i$-dropouts.  Combining these
estimates of halo mass with the observed 
luminosities allow for an estimate
of a mass-to-light ratio,
which can constrain the
connection between the galaxies' observed LF
and their clustering \citep[][]{yang2003a,conroy2006a}.
Numerous effects can alter the
correspondence between bias and halo mass,
and we therefore
view the presented method as primarily a measure
of bias and interpret the halo 
mass estimates with caution.
Our simulations assume that the high-redshift
galaxy population is essentially volume limited
and pure. 
Methods to handle incompleteness for
LBG samples have already been engineered
\citep[e.g.,][]{bouwens2008a,reddy2009a}.
Contamination at a fractional level of $f_{c}$
will lower the bias by an amount $\Deltab\approx(1-f_{c})$
\citep[e.g.,][]{ouchi2004a}. 
Using simulations that include red $z\sim2$ galaxy
interlopers with $f_{c}=0.2$ and a bias of $b\approx5.5$ \citep{quadri2007a},
we find that the measured bias at $z\sim6$ is lowered by $\sim17\%$.
This systematic error is comparable to the statistical error 
on the bias when $\Nfields\gtrsim200$.
Lastly, not every parallel field will be useful for measuring
high-redshift galaxy counts owing, e.g., to possible bright star or Galactic
reddening contamination, and the efficiency of the presented method 
may be correspondingly decreased.

\acknowledgements
I thank the anonymous referee for constructive suggestions,
as well as Richard Ellis, Chuck Steidel, Masami Ouchi, and 
Haojing Yan for helpful discussions.
I am supported by a Hubble Fellowship grant, program number 
HST-HF-51262.01-A provided by NASA from the Space Telescope 
Science Institute, which is operated by the Association of 
Universities for Research in Astronomy, Incorporated, 
under NASA contract NAS5-26555.  
\\


\begin{thebibliography}{50}
\expandafter\ifx\csname natexlab\endcsname\relax\def\natexlab#1{#1}\fi

\bibitem[{{Adelberger} {et~al.}(1998){Adelberger}, {Steidel}, {Giavalisco},
  {Dickinson}, {Pettini}, \& {Kellogg}}]{adelberger1998a}
{Adelberger}, K.~L., {Steidel}, C.~C., {Giavalisco}, M., {Dickinson}, M.,
  {Pettini}, M., \& {Kellogg}, M. 1998, \apj, 505, 18

\bibitem[{{Adelberger} {et~al.}(2005){Adelberger}, {Steidel}, {Pettini},
  {Shapley}, {Reddy}, \& {Erb}}]{adelberger2005a}
{Adelberger}, K.~L., {Steidel}, C.~C., {Pettini}, M., {Shapley}, A.~E.,
  {Reddy}, N.~A., \& {Erb}, D.~K. 2005, \apj, 619, 697

\bibitem[{{Allen} {et~al.}(2005){Allen}, {Moustakas}, {Dalton}, {MacDonald},
  {Blake}, {Clewley}, {Heymans}, \& {Wegner}}]{allen2005a}
{Allen}, P.~D., {Moustakas}, L.~A., {Dalton}, G., {MacDonald}, E., {Blake}, C.,
  {Clewley}, L., {Heymans}, C., \& {Wegner}, G. 2005, \mnras, 360, 1244

\bibitem[{{Arnouts} {et~al.}(1999){Arnouts}, {Cristiani}, {Moscardini},
  {Matarrese}, {Lucchin}, {Fontana}, \& {Giallongo}}]{arnouts1999a}
{Arnouts}, S., {Cristiani}, S., {Moscardini}, L., {Matarrese}, S., {Lucchin},
  F., {Fontana}, A., \& {Giallongo}, E. 1999, \mnras, 310, 540

\bibitem[{{Beckwith} {et~al.}(2006){Beckwith}, {Stiavelli}, {Koekemoer},
  {Caldwell}, {Ferguson}, {Hook}, {Lucas}, {Bergeron}, {Corbin}, {Jogee},
  {Panagia}, {Robberto}, {Royle}, {Somerville}, \& {Sosey}}]{beckwith2006a}
{Beckwith}, S.~V.~W., {et~al.} 2006, \aj, 132, 1729

\bibitem[{{Bouwens} {et~al.}(2006){Bouwens}, {Illingworth}, {Blakeslee}, \&
  {Franx}}]{bouwens2006a}
{Bouwens}, R.~J., {Illingworth}, G.~D., {Blakeslee}, J.~P., \& {Franx}, M.
  2006, \apj, 653, 53

\bibitem[{{Bouwens} {et~al.}(2007){Bouwens}, {Illingworth}, {Franx}, \&
  {Ford}}]{bouwens2007a}
{Bouwens}, R.~J., {Illingworth}, G.~D., {Franx}, M., \& {Ford}, H. 2007, \apj,
  670, 928

\bibitem[{{Bouwens} {et~al.}(2008){Bouwens}, {Illingworth}, {Franx}, \&
  {Ford}}]{bouwens2008a}
---. 2008, \apj, 686, 230

\bibitem[{{Coles} \& {Jones}(1991)}]{coles1991a}
{Coles}, P., \& {Jones}, B. 1991, \mnras, 248, 1

\bibitem[{{Conroy} {et~al.}(2006){Conroy}, {Wechsler}, \&
  {Kravtsov}}]{conroy2006a}
{Conroy}, C., {Wechsler}, R.~H., \& {Kravtsov}, A.~V. 2006, \apj, 647, 201

\bibitem[{{Eisenstein} \& {Hu}(1998)}]{eisenstein1998a}
{Eisenstein}, D.~J., \& {Hu}, W. 1998, \apj, 496, 605

\bibitem[{{Foucaud} {et~al.}(2003){Foucaud}, {McCracken}, {Le F{\`e}vre},
  {Arnouts}, {Brodwin}, {Lilly}, {Crampton}, \& {Mellier}}]{foucaud2003a}
{Foucaud}, S., {McCracken}, H.~J., {Le F{\`e}vre}, O., {Arnouts}, S.,
  {Brodwin}, M., {Lilly}, S.~J., {Crampton}, D., \& {Mellier}, Y. 2003, \aap,
  409, 835

\bibitem[{{Gawiser} {et~al.}(2007){Gawiser}, {Francke}, {Lai}, {Schawinski},
  {Gronwall}, {Ciardullo}, {Quadri}, {Orsi}, {Barrientos}, {Blanc}, {Fazio},
  {Feldmeier}, {Huang}, {Infante}, {Lira}, {Padilla}, {Taylor}, {Treister},
  {Urry}, {van Dokkum}, \& {Virani}}]{gawiser2007a}
{Gawiser}, E., {et~al.} 2007, \apj, 671, 278

\bibitem[{{Giavalisco} {et~al.}(1998){Giavalisco}, {Steidel}, {Adelberger},
  {Dickinson}, {Pettini}, \& {Kellogg}}]{giavalisco1998a}
{Giavalisco}, M., {Steidel}, C.~C., {Adelberger}, K.~L., {Dickinson}, M.~E.,
  {Pettini}, M., \& {Kellogg}, M. 1998, \apj, 503, 543

\bibitem[{{Hamana} {et~al.}(2004){Hamana}, {Ouchi}, {Shimasaku}, {Kayo}, \&
  {Suto}}]{hamana2004a}
{Hamana}, T., {Ouchi}, M., {Shimasaku}, K., {Kayo}, I., \& {Suto}, Y. 2004,
  \mnras, 347, 813

\bibitem[{{Hu} \& {Cohn}(2006)}]{hu2006a}
{Hu}, W., \& {Cohn}, J.~D. 2006, \prd, 73, 067301

\bibitem[{{Kashikawa} {et~al.}(2006){Kashikawa}, {Yoshida}, {Shimasaku},
  {Nagashima}, {Yahagi}, {Ouchi}, {Matsuda}, {Malkan}, {Doi}, {Iye}, {Ajiki},
  {Akiyama}, {Ando}, {Aoki}, {Furusawa}, {Hayashino}, {Iwamuro}, {Karoji},
  {Kobayashi}, {Kodaira}, {Kodama}, {Komiyama}, {Miyazaki}, {Mizumoto},
  {Morokuma}, {Motohara}, {Murayama}, {Nagao}, {Nariai}, {Ohta}, {Okamura},
  {Sasaki}, {Sato}, {Sekiguchi}, {Shioya}, {Tamura}, {Taniguchi}, {Umemura},
  {Yamada}, \& {Yasuda}}]{kashikawa2006a}
{Kashikawa}, N., {et~al.} 2006, \apj, 637, 631

\bibitem[{{Kofman} {et~al.}(1994){Kofman}, {Bertschinger}, {Gelb}, {Nusser}, \&
  {Dekel}}]{kofman1994a}
{Kofman}, L., {Bertschinger}, E., {Gelb}, J.~M., {Nusser}, A., \& {Dekel}, A.
  1994, \apj, 420, 44

\bibitem[{{Komatsu} {et~al.}(2009){Komatsu}, {Dunkley}, {Nolta}, {Bennett},
  {Gold}, {Hinshaw}, {Jarosik}, {Larson}, {Limon}, {Page}, {Spergel},
  {Halpern}, {Hill}, {Kogut}, {Meyer}, {Tucker}, {Weiland}, {Wollack}, \&
  {Wright}}]{komatsu2009a}
{Komatsu}, E., {et~al.} 2009, \apjs, 180, 330

\bibitem[{{Kova{\v c}} {et~al.}(2007){Kova{\v c}}, {Somerville}, {Rhoads},
  {Malhotra}, \& {Wang}}]{kovac2007a}
{Kova{\v c}}, K., {Somerville}, R.~S., {Rhoads}, J.~E., {Malhotra}, S., \&
  {Wang}, J. 2007, \apj, 668, 15

\bibitem[{{Lee} {et~al.}(2009){Lee}, {Giavalisco}, {Conroy}, {Wechsler},
  {Ferguson}, {Somerville}, {Dickinson}, \& {Urry}}]{lee2009a}
{Lee}, K., {Giavalisco}, M., {Conroy}, C., {Wechsler}, R.~H., {Ferguson},
  H.~C., {Somerville}, R.~S., {Dickinson}, M.~E., \& {Urry}, C.~M. 2009, \apj,
  695, 368

\bibitem[{{Lee} {et~al.}(2006){Lee}, {Giavalisco}, {Gnedin}, {Somerville},
  {Ferguson}, {Dickinson}, \& {Ouchi}}]{lee2006a}
{Lee}, K., {Giavalisco}, M., {Gnedin}, O.~Y., {Somerville}, R.~S., {Ferguson},
  H.~C., {Dickinson}, M., \& {Ouchi}, M. 2006, \apj, 642, 63

\bibitem[{{McLure} {et~al.}(2009){McLure}, {Cirasuolo}, {Dunlop}, {Foucaud}, \&
  {Almaini}}]{mclure2009a}
{McLure}, R.~J., {Cirasuolo}, M., {Dunlop}, J.~S., {Foucaud}, S., \& {Almaini},
  O. 2009, \mnras, 395, 2196

\bibitem[{{McLure} {et~al.}(2010){McLure}, {Dunlop}, {Cirasuolo}, {Koekemoer},
  {Sabbi}, {Stark}, {Targett}, \& {Ellis}}]{mclure2010a}
{McLure}, R.~J., {Dunlop}, J.~S., {Cirasuolo}, M., {Koekemoer}, A.~M., {Sabbi},
  E., {Stark}, D.~P., {Targett}, T.~A., \& {Ellis}, R.~S. 2010, \mnras, 403,
  960

\bibitem[{{Mu{\~n}oz} {et~al.}(2010){Mu{\~n}oz}, {Trac}, \&
  {Loeb}}]{munoz2010a}
{Mu{\~n}oz}, J.~A., {Trac}, H., \& {Loeb}, A. 2010, \mnras, 623

\bibitem[{{Norberg} {et~al.}(2001){Norberg}, {Baugh}, {Hawkins}, {Maddox},
  {Peacock}, {Cole}, {Frenk}, {Bland-Hawthorn}, {Bridges}, {Cannon}, {Colless},
  {Collins}, {Couch}, {Dalton}, {De Propris}, {Driver}, {Efstathiou}, {Ellis},
  {Glazebrook}, {Jackson}, {Lahav}, {Lewis}, {Lumsden}, {Madgwick}, {Peterson},
  {Sutherland}, \& {Taylor}}]{norberg2001a}
{Norberg}, P., {et~al.} 2001, \mnras, 328, 64

\bibitem[{{Oesch} {et~al.}(2010){Oesch}, {Bouwens}, {Illingworth}, {Carollo},
  {Franx}, {Labb{\'e}}, {Magee}, {Stiavelli}, {Trenti}, \& {van
  Dokkum}}]{oesch2010a}
{Oesch}, P.~A., {et~al.} 2010, \apjl, 709, L16

\bibitem[{{Ouchi} {et~al.}(2005){Ouchi}, {Hamana}, {Shimasaku}, {Yamada},
  {Akiyama}, {Kashikawa}, {Yoshida}, {Aoki}, {Iye}, {Saito}, {Sasaki},
  {Simpson}, \& {Yoshida}}]{ouchi2005a}
{Ouchi}, M., {et~al.} 2005, \apjl, 635, L117

\bibitem[{{Ouchi} {et~al.}(2009){Ouchi}, {Mobasher}, {Shimasaku}, {Ferguson},
  {Fall}, {Ono}, {Kashikawa}, {Morokuma}, {Nakajima}, {Okamura}, {Dickinson},
  {Giavalisco}, \& {Ohta}}]{ouchi2009a}
---. 2009, \apj, 706, 1136

\bibitem[{{Ouchi} {et~al.}(2003){Ouchi}, {Shimasaku}, {Furusawa}, {Miyazaki},
  {Doi}, {Hamabe}, {Hayashino}, {Kimura}, {Kodaira}, {Komiyama}, {Matsuda},
  {Miyazaki}, {Nakata}, {Okamura}, {Sekiguchi}, {Shioya}, {Tamura},
  {Taniguchi}, {Yagi}, \& {Yasuda}}]{ouchi2003a}
---. 2003, \apj, 582, 60

\bibitem[{{Ouchi} {et~al.}(2004){Ouchi}, {Shimasaku}, {Okamura}, {Furusawa},
  {Kashikawa}, {Ota}, {Doi}, {Hamabe}, {Kimura}, {Komiyama}, {Miyazaki},
  {Miyazaki}, {Nakata}, {Sekiguchi}, {Yagi}, \& {Yasuda}}]{ouchi2004a}
---. 2004, \apj, 611, 685

\bibitem[{{Overzier} {et~al.}(2006){Overzier}, {Bouwens}, {Illingworth}, \&
  {Franx}}]{overzier2006a}
{Overzier}, R.~A., {Bouwens}, R.~J., {Illingworth}, G.~D., \& {Franx}, M. 2006,
  \apjl, 648, L5

\bibitem[{{Peebles}(1980)}]{peebles1980a}
{Peebles}, P.~J.~E. 1980, {The large-scale structure of the universe.}
  (Princeton University Press)

\bibitem[{{Quadri} {et~al.}(2007){Quadri}, {van Dokkum}, {Gawiser}, {Franx},
  {Marchesini}, {Lira}, {Rudnick}, {Herrera}, {Maza}, {Kriek}, {Labb{\'e}}, \&
  {Francke}}]{quadri2007a}
{Quadri}, R., {et~al.} 2007, \apj, 654, 138

\bibitem[{{Ratnatunga}(2002)}]{ratnatunga2002a}
{Ratnatunga}, K. 2002, in HST Proposal, 9488--+

\bibitem[{{Reddy} \& {Steidel}(2009)}]{reddy2009a}
{Reddy}, N.~A., \& {Steidel}, C.~C. 2009, \apj, 692, 778

\bibitem[{{Robertson}(2010)}]{robertson2010a}
{Robertson}, B.~E. 2010, \apj, 713, 1266

\bibitem[{{Schechter}(1976)}]{schechter1976a}
{Schechter}, P. 1976, \apj, 203, 297

\bibitem[{{Sparks}(2002)}]{sparks2002a}
{Sparks}, W. 2002, in HST Proposal, 9575--+

\bibitem[{{Stark} {et~al.}(2007){Stark}, {Loeb}, \& {Ellis}}]{stark2007a}
{Stark}, D.~P., {Loeb}, A., \& {Ellis}, R.~S. 2007, \apj, 668, 627

\bibitem[{{Tinker} {et~al.}(2008){Tinker}, {Kravtsov}, {Klypin}, {Abazajian},
  {Warren}, {Yepes}, {Gottl{\"o}ber}, \& {Holz}}]{tinker2008a}
{Tinker}, J., {Kravtsov}, A.~V., {Klypin}, A., {Abazajian}, K., {Warren}, M.,
  {Yepes}, G., {Gottl{\"o}ber}, S., \& {Holz}, D.~E. 2008, \apj, 688, 709

\bibitem[{{Tinker} {et~al.}(2010){Tinker}, {Robertson}, {Kravtsov}, {Klypin},
  {Warren}, {Yepes}, \& {Gottlober}}]{tinker2010a}
{Tinker}, J.~L., {Robertson}, B.~E., {Kravtsov}, A.~V., {Klypin}, A., {Warren},
  M.~S., {Yepes}, G., \& {Gottlober}, S. 2010, arXiv:1001.3162

\bibitem[{{Trenti}(2008)}]{trenti2008b}
{Trenti}, M. 2008, in HST Proposal, 11700--+

\bibitem[{{Trenti} \& {Stiavelli}(2008)}]{trenti2008a}
{Trenti}, M., \& {Stiavelli}, M. 2008, \apj, 676, 767

\bibitem[{{Vale} \& {Ostriker}(2004)}]{vale2004a}
{Vale}, A., \& {Ostriker}, J.~P. 2004, \mnras, 353, 189

\bibitem[{{van den Bosch} {et~al.}(2003){van den Bosch}, {Yang}, \&
  {Mo}}]{van_den_bosch2003a}
{van den Bosch}, F.~C., {Yang}, X., \& {Mo}, H.~J. 2003, \mnras, 340, 771

\bibitem[{{Yan}(2008)}]{yan2008a}
{Yan}, H. 2008, in HST Proposal, 11702--+

\bibitem[{{Yan} {et~al.}(2009){Yan}, {Windhorst}, {Hathi}, {Cohen}, {Ryan},
  {O'Connell}, \& {McCarthy}}]{yan2009a}
{Yan}, H., {Windhorst}, R., {Hathi}, N., {Cohen}, S., {Ryan}, R., {O'Connell},
  R., \& {McCarthy}, P. 2009, arXiv:0910.0077

\bibitem[{{Yang} {et~al.}(2003){Yang}, {Mo}, \& {van den Bosch}}]{yang2003a}
{Yang}, X., {Mo}, H.~J., \& {van den Bosch}, F.~C. 2003, \mnras, 339, 1057

\bibitem[{{Zehavi} {et~al.}(2002){Zehavi}, {Blanton}, {Frieman}, {Weinberg},
  {Mo}, {Strauss}, {Anderson}, {Annis}, {Bahcall}, {Bernardi}, {Briggs},
  {Brinkmann}, {Burles}, {Carey}, {Castander}, {Connolly}, {Csabai},
  {Dalcanton}, {Dodelson}, {Doi}, {Eisenstein}, {Evans}, {Finkbeiner},
  {Friedman}, {Fukugita}, {Gunn}, {Hennessy}, {Hindsley}, {Ivezi{\'c}}, {Kent},
  {Knapp}, {Kron}, {Kunszt}, {Lamb}, {Leger}, {Long}, {Loveday}, {Lupton},
  {McKay}, {Meiksin}, {Merrelli}, {Munn}, {Narayanan}, {Newcomb}, {Nichol},
  {Owen}, {Peoples}, {Pope}, {Rockosi}, {Schlegel}, {Schneider}, {Scoccimarro},
  {Sheth}, {Siegmund}, {Smee}, {Snir}, {Stebbins}, {Stoughton}, {SubbaRao},
  {Szalay}, {Szapudi}, {Tegmark}, {Tucker}, {Uomoto}, {Vanden Berk}, {Vogeley},
  {Waddell}, {Yanny}, \& {York}}]{zehavi2002a}
{Zehavi}, I., {et~al.} 2002, \apj, 571, 172

\end{thebibliography}
\end{document}